\documentclass[pre]{revtex4-2}

\usepackage{amsmath, amssymb, mathtools, siunitx, physics}
\usepackage[dvipdfmx]{graphicx}

\DeclareMathOperator{\sgn}{sgn}

\begin{document}
\title{Exact solution of the propagation of ON--OFF signals by dispersive waves}

\author{Ken Yamamoto}
\email{yamamot@sci.u-ryukyu.ac.jp}

\affiliation{Faculty of Science, University of the Ryukyus, Senbaru 1, Nishihara, Okinawa 903--0213, Japan}

\begin{abstract}
The propagation of ON--OFF signals with dispersive waves is examined in this study.
An integral-form exact solution for a simple ON--OFF switching event is derived, which holds for any dispersion relation.
The integral can be exactly calculated for two types of dispersion relations.
Further, the analysis of these solutions shows that the ON--OFF signal propagates with the group velocity and that the boundary thickness of the signal increases with time, typically at a rate proportional to the square root of time, owing to dispersion.
Additionally, an approximate solution for a general dispersion relation is derived, and a for a higher-complexity ON--OFF switching pattern is constructed.
\end{abstract}

\maketitle

keywords: wave, dispersion relation, group velocity, exact solution, Fresnel function

\section{Introduction}
Waves appear in various physical systems~\cite{Crawford, Lighthill, Nettel, Narayanan}, spanning quantum~\cite{Phillips} to astronomical~\cite{Creighton} scales.
The control and detection of waves have application prospects in a variety of fields, such as wireless communications~\cite{Tse}, acoustics~\cite{Pierce}, architectonics~\cite{Taranath}, and meteorology~\cite{Holton}.

The dispersion of waves means that the phase velocity $v_\mathrm{p}=\omega(k)/k$ and the group velocity $v_\mathrm{g}=d\omega(k)/dk$ are different, where $\omega(k)$ is the dispersion relation, i.e., angular frequency $\omega$ as a function of wavenumber $k$.
The existence of two kinds of wave velocities causes complex, sometimes seemingly paradoxical phenomena, such as fast (or slow) light~\cite{Boyd} whose group velocity is faster (or much slower) than the speed of light in vacuum, the Kelvin wake created by a moving ship on the water surface~\cite{Lighthill}, and the Rossby wave associated with global climate~\cite{Holton}.

When a wave propagates in a dispersive medium, its shape is generally deformed (distorted and broadened) with time because component sine waves travel at different phase velocities.
Sommerfeld~\cite{Sommerfeld} and Brillouin~\cite{Brillouin} analyzed a light signal (electromagnetic wave) incident to dispersive dielectric and theoretically predicted the appearance of wave packets, called precursors or forerunners, preceding the main signal.
Their calculation was later corrected and modified~\cite{Oughstun}, and precursors consistent with the theory were experimentally observed~\cite{Jeong}.
The Brillouin precursor, containing low-frequency wave components, exhibits power-law decay with propagation distance~\cite{Oughstun2005} and is expected to be applied to communication~\cite{Alejos} and radar~\cite{Cartwright}.

In the calculation of electromagnetic precursors, the asymptotic analysis has been performed after introducing approximations.
The calculation of propagating waves through a dispersive medium is generally difficult, because complicated Fourier-type integrals are involved.
Such integrals cannot be calculated exactly using the existing mathematical functions, and approximation and asymptotic method have been necessarily employed to evaluate the integrals.

In this study, the exact solution of the propagation of signal waves in dispersive media is investigated.
A simple ON--OFF switching event is analyzed, in which the wave source does not oscillate in time $t<0$, and harmonic oscillation is initiated at $t=0$.
The exact solution is derived for the general dispersion relation $\omega=\omega(k)$ in terms of an integral representing a superposition of sine waves.
It is shown that the integral can be exactly calculated to obtain the closed form for dispersion relations $\omega=ck$ and $\omega=Dk^2$.
The properties of these closed-form solutions are explained.
Notably, from the solution for $\omega=Dk^2$, it is determined that the boundary between the ON and OFF states travels at group velocity $v_\mathrm{g}$ and that the boundary thickness increases at a rate proportional to $t^{1/2}$ with time.
Furthermore, an approximate solution for a general dispersion relation is derived, and a solution for a higher-complexity ON--OFF switching pattern is constructed.

\section{Formulation of problem and integral-form solution}\label{sec2}
To investigate the simplest situation, it is assumed that the wave is expressed by a real-valued function, $u(x,t)$, which satisfies a \emph{linear} partial differential equation with a \emph{second-order derivative in time}.
The wave equation shown in Eq.~\eqref{eq:waveeq} is a typical example of such an equation.

The analysis begins from the situation of switching from the OFF to the ON state at $t=0$.
Specifically, at $t=0$, the wave source at $x=0$ initiates harmonic oscillation with amplitude $A$ and angular frequency $\Omega$, and the wave propagates in the $x>0$ region.
This incident signal is the same as Sommerfeld and Brillouin used~\cite{Brillouin}.
Let $K$ denote the wavenumber such that $\Omega=\omega(K)$, where $\omega=\omega(k)$ is the dispersion relation.

Evidently, $u(x,t)=0$ for any $x\ge0$ at any $t<0$.
To construct the solution for $t\ge0$, the initial condition for $u(x,t)$ is expressed at $t=0$.
Considering the assumption of the second-order derivative in time, the initial condition corresponding to the OFF state before $t=0$ is given by
\begin{align}
&u(x, 0)=0,\label{eq:initial1}\\
&\frac{\partial u}{\partial t}(x, 0)=0\label{eq:initial2}
\end{align}
for any $x>0$.
In addition, the boundary condition for the wave source at $x=0$ becomes
\begin{equation}
u(0, t)=A\sin\Omega t
\label{eq:boundary}
\end{equation}
for $t\ge0$.

Since linearity is assumed, the superposition principle is valid for the wave in question.
The solution $u(x,t)$ which satisfies conditions~\eqref{eq:initial1}--\eqref{eq:boundary} can be constructed by a superposition of sine waves $\sin(\omega(k)t\pm kx)$ and $\cos(\omega(k)t\pm kx)$ as follows.
To begin with, the following superposition,
\begin{equation}
u(x, t)=\frac{A}{2}\sin(\Omega t-Kx)+\frac{A}{2}\sin(\Omega t+Kx)
-\frac{A}{2}\Omega\int_0^\infty \varphi(k)[\cos(\omega(k)t-kx)-\cos(\omega(k)t+kx)]dk
\label{eq:solution_int}
\end{equation}
satisfies Eqs.~\eqref{eq:initial1} and \eqref{eq:boundary} for any weight function $\varphi(k)$.
The function $\varphi(k)$ is determined so that condition~\eqref{eq:initial2} is satisfied.
Substituting Eq.~\eqref{eq:solution_int} into \eqref{eq:initial2},
\[
\int_0^\infty \varphi(k)\omega(k)\sin(kx) dk=\cos(Kx)
\]
for any $x>0$.
As $x>0$, this expression can be rewritten as
\begin{equation}
\int_0^\infty \varphi(k)\omega(k)\sin(kx)dk=\cos(Kx)\sgn(x),
\label{eq:cos_sgn}
\end{equation}
where $\sgn$ is the signum function defined by
\[
\sgn(x)=
\begin{dcases}
1 & x>0,\\
0 & x=0,\\
-1 & x<0.
\end{dcases}
\]
The formula
\[
\int_0^\infty \frac{\sin(kx)}{k}dk=\frac{\pi}{2}
\]
for $x>0$ is well known in complex calculus~\cite{Ahlfors}, which is sometimes called the Dirichlet integral~\cite{Bartle}.
By extending this relation to $x=0$ and $x<0$,
\begin{equation}
\sgn(x)=\frac{2}{\pi}\int_0^\infty \frac{1}{k}\sin(kx)dk.
\label{eq:sgn}
\end{equation}
Therefore,
\begin{equation}
\cos(Kx)\sgn(x)=\frac{2}{\pi}\pv{\int_0^\infty \frac{k}{k^2-K^2}\sin(kx)dk},
\label{eq:cos_sgn2}
\end{equation}
where $\pv$ represents the Cauchy principal value at the singular point $k=K$ of the integrand.
The derivation of Eq.~\eqref{eq:cos_sgn2} is presented in Appendix~\ref{apdx:a2}.
By the definition of the Cauchy principal-value integral, Eq.~\eqref{eq:cos_sgn2} means that
\[
\cos(Kx)\sgn(x)=\frac{2}{\pi}\lim_{\varepsilon\to0+}\left[\int_0^{K-\varepsilon} \frac{k}{k^2-K^2}\sin(kx)dk+\int_{K+\varepsilon}^\infty \frac{k}{k^2-K^2}\sin(kx)dk\right].
\]
Equations~\eqref{eq:cos_sgn} and \eqref{eq:cos_sgn2} yield
\[
\varphi(k)=\frac{2}{\pi}\frac{k}{k^2-K^2}\frac{1}{\omega(k)},
\]
and thus,
\begin{equation}
u(x, t)=
\begin{dcases}
0 & (t<0),\\
\dfrac{A}{2}\sin(\Omega t-Kx)+\dfrac{A}{2}\sin(\Omega t+Kx)
-\dfrac{A}{\pi}\Omega\pv{\displaystyle\int_0^\infty \dfrac{k}{k^2-K^2}\dfrac{1}{\omega(k)}[\cos(\omega(k)t-kx)-\cos(\omega(k)t+kx)]dk} & (t\ge0).
\end{dcases}
\label{eq:general_solution}
\end{equation}
This is the integral-form solution to the initial and boundary value problem [Eqs.~\eqref{eq:initial1}--\eqref{eq:boundary}] valid for the general dispersion relation $\omega(k)$.

The exact calculation of the integral in Eq.~\eqref{eq:general_solution} is feasible only for a simple $\omega(k)$ because $\omega(k)$ appears in the denominator and the arguments of the cosine functions.
In the following sections, the exact closed forms of $u(x, t)$ for two specific dispersion relations are derived, and the corresponding wave propagation is discussed.

\section{Nondispersive case: $\omega(k)=ck$}\label{sec3}
The simplest dispersion relation, having no dispersion, is $\omega(k)=ck$, with a constant $c>0$.
The wave equation
\begin{equation}
\frac{\partial^2 u}{\partial t^2}=c^2\frac{\partial^2 u}{\partial x^2}
\label{eq:waveeq}
\end{equation}
has this dispersion relation.
This dispersion relation is found in electromagnetic waves in vacuum, sound waves, shallow-water waves, and gravitational waves~\cite{Thorne}.
Both the phase velocity $v_\mathrm{p}$ and group velocity $v_\mathrm{g}$ become $v_\mathrm{p}=v_\mathrm{g}=c$ in this case.
The integral term in Eq.~\eqref{eq:general_solution} becomes
\[
-\frac{A\Omega}{\pi c}\pv{\int_0^\infty \frac{1}{k^2-K^2}[\cos((ct-x)k)-\cos((ct+k))]dk}.
\]
Using the formula (see Appendix~\ref{apdx:a3})
\[
\pv{\int_0^\infty \frac{1}{k^2-K^2}\cos(\alpha k)dk} = -\frac{\pi\sin(\alpha K)}{2K}\sgn(\alpha)
\]
for $\alpha=ct-x$ and $ct+x$ [and noting that $\sgn(ct+x)=1$ for any $x>0$ and $t\ge0$], solution~\eqref{eq:general_solution} becomes
\begin{equation}
u(x, t)=\frac{A}{2}(1+\sgn(ct-x))\sin(\Omega t-Kx)
=\begin{dcases}
A\sin(\Omega t-Kx) & (x\le ct),\\
0 & (x>ct).
\end{dcases}
\label{eq:solution_ck}
\end{equation}
The wave travels with velocity $c$, maintaining its sinusoidal shape.
This solution is a function of the single variable $x-ct[=(Kx-\Omega t)/K]$, which is an example of d'Alembert's solution for the wave equation.
Strictly, this solution is not a classical solution but a weak solution to the wave equation~\cite{Cain}, because it is not smooth at $x=ct$.

\begin{figure}[t!]\centering
\raisebox{33mm}{(a)}
\includegraphics[scale=0.7]{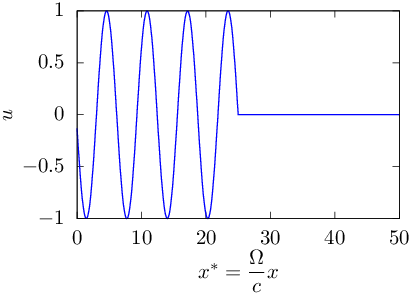}
\hspace{5mm}
\raisebox{33mm}{(b)}
\includegraphics[scale=0.7]{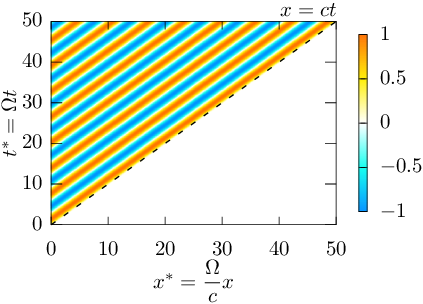}
\caption{
Wave propagation for the nondispersive $\omega(k)=ck$.
The dimensionless quantities, $x^\ast=(\Omega/c)x$, $t^\ast=\Omega t$, and $A=1$, are used.
(a) Snapshot of $u(x, t)$ at $t^\ast=25$.
(b) Spatiotemporal diagram of $u$.
The dashed diagonal line indicates $x=ct$.
}
\label{fig1}
\end{figure}

Figure~\ref{fig1} illustrates the wave propagation in terms of dimensionless space $x^\ast=(\Omega/c)x$ and time $t^\ast=\Omega t$, with dimensionless amplitude $A=1$.
Figure~\ref{fig1}(a) shows the snapshot of $u(x,t)$ at $t^\ast=25$.
The ON state (oscillation) and OFF state ($u=0$) are sharply separated at $x^\ast=25(=t^\ast)$.
The spatiotemporal structure of $u(x,t)$ is illustrated in Fig.~\ref{fig1}(b).
The dashed diagonal line ($x^\ast=t^\ast$ or equivalently $x=ct$) represents the boundary between the ON and OFF states, indicating that the ON--OFF signal travels with velocity $c$.

\section{Dispersive example: $\omega(k)=Dk^2$}\label{sec4}
\subsection{Exact solution}
In this section, the exact solution of $u(x, t)$ for a dispersive case of $\omega(k)=Dk^2$ ($D>0$ is a constant) is provided.
The phase velocity is $v_\mathrm{p}=Dk$, and the group velocity is $v_\mathrm{g}=2Dk$.
Thus, $v_\mathrm{g}=2v_\mathrm{p}$ for any $k$.
This dispersion relation arises in the uniform Euler--Bernoulli beam~\cite{Hagedorn} with no applied load, obeying
\begin{equation}
\mu\frac{\partial^2 u}{\partial t^2}=-EI\frac{\partial^4 u}{\partial x^4},
\label{eq:EulerBernoulli}
\end{equation}
where $u$ is the deflection, $\mu$ is the mass per unit length, and $EI$ is the flexural rigidity of the beam.
The relation $D^2=EI/\mu$ is easily obtained.
The Schr\"odinger equation for a free particle shares this dispersion relation~\cite{Phillips}, although the wave function is complex-valued.

In this case, Eq.~\eqref{eq:general_solution} for $t\ge0$ becomes
\[
u(x, t)=\frac{A}{2}\sin(\Omega t-Kx)+\frac{A}{2}\sin(\Omega t+Kx)
-\frac{A\Omega}{\pi D}\pv{\int_0^\infty \frac{1}{k^2-K^2}\frac{1}{k}[\cos(Dk^2t-kx)-\cos(Dk^2t+kx)]dk}.
\]
As explained in Appendix~\ref{apdx:b}, the integral can be exactly calculated, and the solution becomes
\begin{align}
u(x, t)&=\frac{A}{2}\left[1+C\left(\frac{2DKt-x}{\sqrt{2\pi Dt}}\right)+S\left(\frac{2DKt-x}{\sqrt{2\pi Dt}}\right)\right]\sin(\Omega t-Kx)\nonumber\\
&\quad+\frac{A}{2}\left[C\left(\frac{2DKt-x}{\sqrt{2\pi Dt}}\right)-S\left(\frac{2DKt-x}{\sqrt{2\pi Dt}}\right)\right]\cos(\Omega t-Kx)\nonumber\\
&\quad+\frac{A}{2}\left[1-C\left(\frac{2DKt+x}{\sqrt{2\pi Dt}}\right)-S\left(\frac{2DKt+x}{\sqrt{2\pi Dt}}\right)\right]\sin(\Omega t+Kx)\nonumber\\
&\quad-\frac{A}{2}\left[C\left(\frac{2DKt+x}{\sqrt{2\pi Dt}}\right)-S\left(\frac{2DKt+x}{\sqrt{2\pi Dt}}\right)\right]\cos(\Omega t+Kx)\nonumber\\
&\quad+A\left[C\left(\frac{x}{\sqrt{2\pi Dt}}\right)-S\left(\frac{x}{\sqrt{2\pi Dt}}\right)\right],
\label{eq:solution_Dk2}
\end{align}
where $S$ and $C$ are the Fresnel functions~\cite{Olver}, defined as
\[
S(x)=\int_0^x \sin\left(\frac{\pi}{2}s^2\right)ds,\quad
C(x)=\int_0^x \cos\left(\frac{\pi}{2}s^2\right)ds,
\]
respectively.
This result is heavily complicated; however the exact solution is indeed obtained in a closed form.

\begin{figure}[t!]\centering
\raisebox{33mm}{(a)}
\includegraphics[scale=0.7]{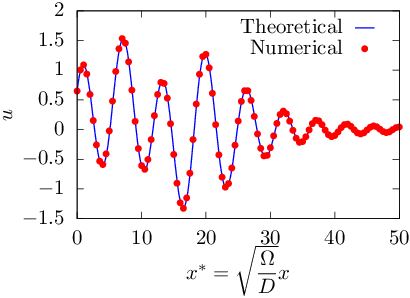}
\hspace{5mm}
\raisebox{33mm}{(b)}
\includegraphics[scale=0.7]{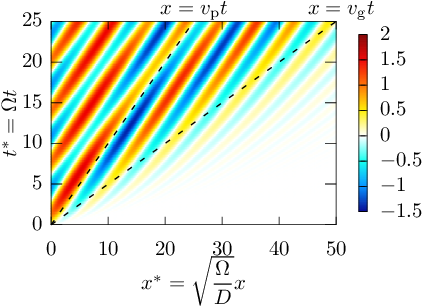}
\caption{
Wave propagation for $\omega(k)=Dk^2$.
The dimensionless quantities, $x^\ast=\sqrt{\Omega/D}x$, $t^\ast=\Omega t$, and $A=1$, are used.
(a) Snapshot of $u(x, t)$ at $t^\ast=15$.
The solid curve shows the exact solution~\eqref{eq:solution_Dk2}, and the circles represent a numerical result obtained using the finite element method.
(b) Spatiotemporal diagram of $u$.
The two dashed lines represent $x=v_\mathrm{p}t$ and $x=v_\mathrm{g}t$.
}
\label{fig2}
\end{figure}

The properties of solution~\eqref{eq:solution_Dk2} are presented in Fig.~\ref{fig2}.
In this case, the dimensionless space and time are given by $x^\ast=\sqrt{\Omega/D}x$ and $t^\ast=\Omega t$, respectively.
The solid curve in Fig.~\ref{fig2}(a) shows the snapshot of $u(x, t)$ at $t^\ast=15$.
Unlike the case in Fig.~\ref{fig1}(a), the wave in small $x^\ast$ is not perfectly periodic, and a small-amplitude wave appears to leak out into the large-$x^\ast$ region.
That is, the boundary between the ON and OFF states is not sharp, compared with that in the case of $\omega(k)=ck$.
Furthermore, some $x$ points, e.g., $x^\ast\approx7$ in Fig.~\ref{fig2}(a), take the value of $|u(x,t)|>1=A$, which is greater than the oscillation amplitude at the wave source.
The circles in Fig.~\ref{fig2}(a) show a numerical results obtained with the finite element method~\cite{Kikuchi} for the Euler--Bernoulli beam~\eqref{eq:EulerBernoulli}, which support the validity of the exact solution~\eqref{eq:solution_Dk2}.

Figure~\ref{fig2}(b) shows the spatiotemporal diagram of $u(x,t)$.
Compared with Fig.~\ref{fig1}(b) for the nondispersive case, this graph appears to be disordered.
The dashed lines indicate $x=v_\mathrm{p}t$ (equivalently $x^\ast=t^\ast$) and $x=v_\mathrm{g}t$ ($x^\ast=2t^\ast$).
The group velocity $v_\mathrm{g}$ is more appropriate than the phase velocity $v_\mathrm{p}$ for the moving velocity of the ON--OFF boundary.
Further analysis is performed in the following subsection.

\subsection{Signal propagation for $\omega(k)=Dk^2$}
The exact solution~\eqref{eq:solution_Dk2} has a complicated form.
Here, further properties for this solution are derived, particularly regarding signal propagation.

The Fresnel functions have limit values $C(\pm\infty)=S(\pm\infty)=\pm1/2$, known as the Fresnel integral~\cite{Ahlfors, Olver}.
Therefore, for a sufficiently large $t$ such that $x\ll 2DKt=v_\mathrm{g}t$, the exact solution~\eqref{eq:solution_Dk2} is approximated to $u(x,t)\simeq A\sin(\Omega t-Kx)$.
That is, within the large-$t$ limit, $u(x,t)$ becomes a sine wave with amplitude $A$ and angular frequency $\Omega$, equal to those of the wave source.

\begin{figure}\centering
\includegraphics[scale=0.7]{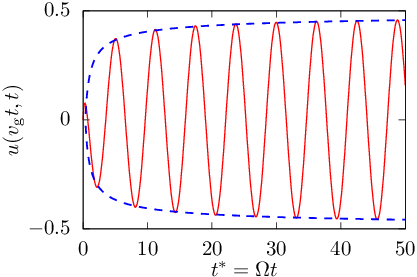}
\caption{
Oscillation of $u(2DKt, t)=u(v_\mathrm{g}t, t)$, given by Eq.~\eqref{eq:oscillate_vg} (solid curve), and the asymptotic amplitude~\eqref{eq:approx_amplitude} (dashed curve).
}
\label{fig3}
\end{figure}

From Fig.~\ref{fig2}(b), the ON--OFF signal is suggested to propagate at group velocity $v_\mathrm{g}=2DK$ rather than phase velocity $v_\mathrm{p}=DK$.
Along the point that moves with velocity $v_\mathrm{g}=2DK$, namely $x=v_\mathrm{g}t=2DKt$, the time variation of $u$ is given by
\begin{align}
u(2DKt, t)&=-\frac{A}{2}\sin(\Omega t)
+A\left[C\left(\sqrt{\frac{2\Omega t}{\pi}}\right)-S\left(\sqrt{\frac{2\Omega t}{\pi}}\right)\right]\nonumber\\
&\quad+\frac{A}{2}\left[1-C\left(2\sqrt{\frac{2\Omega t}{\pi}}\right)-S\left(2\sqrt{\frac{2\Omega t}{\pi}}\right)\right]\sin(3\Omega t)\nonumber\\
&\quad-\frac{A}{2}\left[C\left(2\sqrt{\frac{2\Omega t}{\pi}}\right)-S\left(2\sqrt{\frac{2\Omega t}{\pi}}\right)\right]\cos(3\Omega t).
\label{eq:oscillate_vg}
\end{align}
Using the asymptotic expansion of the Fresnel functions~\cite{Olver},
\begin{equation}
C(z)\sim\frac{1}{2}+\frac{1}{\pi z}\sin\left(\frac{\pi}{2}z^2\right),\quad
S(z)\sim\frac{1}{2}-\frac{1}{\pi z}\cos\left(\frac{\pi}{2}z^2\right),
\label{eq:asymptotic}
\end{equation}
the following is obtained:
\[
u(2DKt, t)=-\frac{A}{2}\sin(\Omega t)+\frac{3A}{4\sqrt{2\pi\Omega t}}(\sin(\Omega t)+\cos(\Omega t)).
\]
Since the second term slowly decays with $t^{-1/2}$, $u(2DKt, t)$ oscillates with amplitude $A/2$ for a sufficiently large $t$.
More strictly,
\[
u(2DKt, t)=-\frac{A}{2}\left(1-\frac{3}{2\sqrt{2\pi\Omega t}}\right)\sin(\Omega t)+\frac{3A}{4\sqrt{2\pi\Omega t}}\cos(\Omega t),
\]
whose amplitude becomes
\begin{equation}
\frac{A}{2}\sqrt{\left(1-\frac{3}{2\sqrt{2\pi\Omega t}}\right)^2+\left(\frac{3}{2\sqrt{2\pi\Omega t}}\right)^2}
=\frac{A}{2}\left(1-\frac{3}{2\sqrt{2\pi\Omega t}}+O(t^{-1})\right).
\label{eq:approx_amplitude}
\end{equation}
Thus, the amplitude at $x=v_\mathrm{g}t$ is slightly smaller than $A/2$ for a finite $t$.
In Fig.~\ref{fig3}, $u(2DKt, t)$ is presented in the exact form~\eqref{eq:oscillate_vg} with the solid curve, and the asymptotic amplitude~\eqref{eq:approx_amplitude} up to $O(t^{-1/2})$ terms is presented with the dashed curve.
From this figure, the asymptotic amplitude appears to be close to the true amplitude for $t^\ast\gtrsim10$.

The envelope provides useful information for investigating the signal propagation.
However, the expression of the envelope's shape for the complicated solution~\eqref{eq:solution_Dk2} is difficult to compute.
Instead of exact calculations, an approximate envelope near $x=v_\mathrm{g}t=2DKt$ is derived.

For $x\gg\sqrt{2\pi Dt}$, $u(x,t)$ in Eq.~\eqref{eq:solution_Dk2} near $x=v_\mathrm{g}t=2DKt$ can be approximated to
\begin{align*}
u(x, t)&\simeq\frac{A}{2}\left[1+C\left(\frac{2DKt-x}{\sqrt{2\pi Dt}}\right)+S\left(\frac{2DKt-x}{\sqrt{2\pi Dt}}\right)\right]\sin(\Omega t-Kx)+\frac{A}{2}\left[C\left(\frac{2DKt-x}{\sqrt{2\pi Dt}}\right)-S\left(\frac{2DKt-x}{\sqrt{2\pi Dt}}\right)\right]\cos(\Omega t-Kx)\\
&\eqqcolon\tilde{A}(x,t)\sin(\Omega t-Kx+\tilde{\phi}(x,t)),
\end{align*}
where
\begin{align}
\tilde{A}(x,t)&=\frac{A}{2}\sqrt{\left[1+C\left(\frac{2DKt-x}{\sqrt{2\pi Dt}}\right)+S\left(\frac{2DKt-x}{\sqrt{2\pi Dt}}\right)\right]^2+\left[C\left(\frac{2DKt-x}{\sqrt{2\pi Dt}}\right)-S\left(\frac{2DKt-x}{\sqrt{2\pi Dt}}\right)\right]^2}\nonumber\\
&=\frac{A}{\sqrt{2}}\sqrt{\left[C\left(\frac{2DKt-x}{\sqrt{2\pi Dt}}\right)+\frac{1}{2}\right]^2+\left[S\left(\frac{2DKt-x}{\sqrt{2\pi Dt}}\right)+\frac{1}{2}\right]^2}
\label{eq:approximate_envelope}
\end{align}
and
\[
\tilde{\phi}(x,t)=\arctan\frac{C\left(\frac{2DKt-x}{\sqrt{2\pi Dt}}\right)-S\left(\frac{2DKt-x}{\sqrt{2\pi Dt}}\right)}{1+C\left(\frac{2DKt-x}{\sqrt{2\pi Dt}}\right)+S\left(\frac{2DKt-x}{\sqrt{2\pi Dt}}\right)}.
\]
Hence, the approximate envelope is given by $\tilde{A}(x, t)$.

As a special value, $\tilde{A}(2DKt, t)=A/2$ is obtained for any $t$.
That is, $x=v_\mathrm{g}t$ is exactly the middle point of the ON and OFF states for the approximate envelope $\tilde{A}$.
This property strongly suggests that the ON--OFF signal propagates with the group velocity.

\begin{figure}\centering
\raisebox{34mm}{(a)}
\includegraphics[scale=0.7]{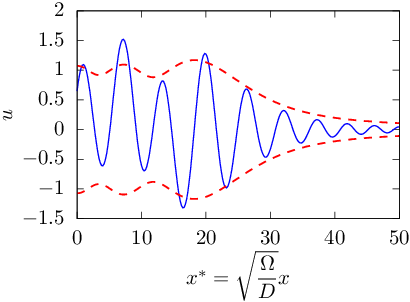}
\hspace{3mm}
\raisebox{34mm}{(b)}
\includegraphics[scale=0.7]{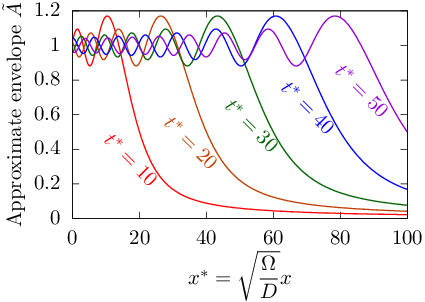}
\hspace{3mm}
\raisebox{34mm}{(c)}
\includegraphics[scale=0.7]{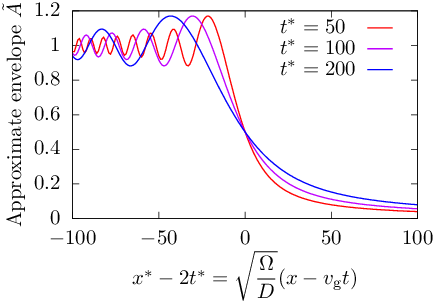}
\caption{
(a) Snapshot of $u$ (solid curve) and the approximate envelope $\tilde{A}(x,t)$, given by Eq.~\eqref{eq:approximate_envelope} (dashed curve) at $t^\ast=15$.
(b) Profile of $\tilde{A}(x, t)$ at $t^\ast=10$, $20$, $30$, $40$, and $50$.
(c) Comparison of $\tilde{A}(x, t)$ at $t^\ast=50$, $100$, and $200$ as a function of $x-v_\mathrm{g}t$.
}
\label{fig4}
\end{figure}

Figure~\ref{fig4} illustrates the properties of the approximate envelope $\tilde{A}(x,t)$.
Figure~\ref{fig4}(a) shows the snapshot of $u$ with the solid curve and $\tilde{A}$ with the dashed curve at $t^\ast=15$.
[The solid curve is the same as the solid curve in Fig.~\ref{fig2}(a).]
Although the envelope generally lies outside the wave, the approximate envelope $\tilde{A}(x,t)$ lies inside of $u(x, t)$ at some $x$ point, e.g., $x^\ast\approx7$ in Fig.~\ref{fig4}(a).
Despite such a shortcoming, the overall shape of $u$ can be broadly described by $\tilde{A}(x,t)$.
Thus, $\tilde{A}(x, t)$ is analyzed as a substitute for the true envelope of $u(x, t)$.
It is expected that $\tilde{A}(x, t)$ becomes very close to the true envelope around $x=2DKt$, with a sufficiently large $t$.

Figure~\ref{fig4}(b) shows the progression of $\tilde{A}(x, t)$ at $t^\ast=10$, $20$, $30$, $40$, and $50$.
To compare the shape of $\tilde{A}(x, t)$ at different $t$ values in more detail, 
Fig.~\ref{fig4}(c) shows $\tilde{A}$ at $t^\ast=50$, $100$, and $200$ using the spatial coordinate $x-v_\mathrm{g}t=x-2DKt$, which moves with group velocity $v_\mathrm{g}$.
From these graphs, it is clear that the envelope $\tilde{A}(x, t)$ moves rightward, with its slope becoming gentler.

More quantitatively, by differentiating Eq.~\eqref{eq:approximate_envelope} with respect to $x$, the slope of $\tilde{A}(x, t)$ at $x=v_\mathrm{g}t$ becomes
\[
\frac{\partial\tilde{A}}{\partial x}(v_\mathrm{g}t, t)=-\frac{A}{\sqrt{2\pi Dt}}.
\]
That is, the slope of the approximate envelope $\tilde{A}(x,t)$ becomes gentler as $t^{-1/2}$ with time.
In other words, the thickness of the transition layer between the ON and OFF states increases as $t^{1/2}$ with time.
This indicates that the ON--OFF signal becomes indistinct as the wave propagates.
The key exponent $1/2$ can also be found by considering that $\tilde{A}$ is a function of the single variable $(x-2DKt)/\sqrt{t}$.
This is different from the nondispersive case where the signal propagates while maintaining distinct ON and OFF regions.

\section{Approximate solution for the general dispersion relation}\label{sec5}
As noted at the end of Section~\ref{sec2}, the exact calculation of the integral in Eq.~\eqref{eq:general_solution} will be performed only for simple dispersion relations $\omega(k)$, such as $\omega(k)=ck$ and $\omega(k)=Dk^2$ in Sections~\ref{sec3} and \ref{sec4}, respectively.
Here, an approximate calculation of the integral for a general dispersion relation $\omega(k)$ is performed.

Using the Taylor expansion of $\omega(k)$ up to $k^2$ terms,
\[
\omega(k\pm K)=\Omega\pm v_\mathrm{g}k+\gamma k^2+O(k^3),
\]
where
\[
\gamma=\frac{1}{2}\left.\frac{d^2\omega(k)}{dk^2}\right|_{k=K},
\]
the following result is obtained:
\begin{align}
u(x,t)&=\frac{A}{2}\left[1+C\left(\frac{v_\mathrm{g}t-x}{\sqrt{2\pi|\gamma|t}}\right)+S\left(\frac{v_\mathrm{g}t-x}{\sqrt{2\pi|\gamma|t}}\right)\right]\sin(\Omega t-Kx)\nonumber\\
&\quad+\frac{A}{2}\sgn(\gamma)\left[C\left(\frac{v_\mathrm{g}t-x}{\sqrt{2\pi|\gamma|t}}\right)-S\left(\frac{v_\mathrm{g}t-x}{\sqrt{2\pi|\gamma|t}}\right)\right]\cos(\Omega t-Kx)\nonumber\\
&\quad+\frac{A}{2}\left[1-C\left(\frac{v_\mathrm{g}t+x}{\sqrt{2\pi|\gamma|t}}\right)-S\left(\frac{v_\mathrm{g}t+x}{\sqrt{2\pi|\gamma|t}}\right)\right]\sin(\Omega t+Kx)\nonumber\\
&\quad-\frac{A}{2}\sgn(\gamma)\left[C\left(\frac{v_\mathrm{g}t+x}{\sqrt{2\pi|\gamma|t}}\right)-S\left(\frac{v_\mathrm{g}t+x}{\sqrt{2\pi|\gamma|t}}\right)\right]\cos(\Omega t+Kx).
\label{eq:approximate}
\end{align}
The calculation process is explained in Appendix~\ref{apdx:c}.
In the derivation, it is assumed that $\omega(k)=\omega(-k)$, i.e., $\omega(k)$ is an even function, which reflects the spatial inversion symmetry.

Here, this approximate solution is compared with the exact solutions obtained in the previous two sections.
For $\omega(k)=ck$, $v_\mathrm{g}=c$ and $\gamma=0$.
Instead of directly substituting $\gamma=0$, the following limit values are adopted:
\[
\lim_{\gamma\to0}C\left(\frac{v_\mathrm{g}t+x}{\sqrt{2\pi|\gamma|t}}\right)=C(\infty)=\frac{1}{2},\quad
\lim_{\gamma\to0}S\left(\frac{v_\mathrm{g}t+x}{\sqrt{2\pi|\gamma|t}}\right)=S(\infty)=\frac{1}{2}\]
for any $x,t>0$, and
\begin{align*}
\lim_{\gamma\to0}C\left(\frac{v_\mathrm{g}t-x}{\sqrt{2\pi|\gamma|t}}\right)&=C(\sgn(v_\mathrm{g}t-x)\infty)=\sgn(v_\mathrm{g}t-x)\frac{1}{2},\\
\lim_{\gamma\to0}S\left(\frac{v_\mathrm{g}t-x}{\sqrt{2\pi|\gamma|t}}\right)&=S(\sgn(v_\mathrm{g}t-x)\infty)=\sgn(v_\mathrm{g}t-x)\frac{1}{2}.
\end{align*}
By substituting these values, the approximate solution~\eqref{eq:approximate} becomes identical to the exact solution~\eqref{eq:solution_ck}.

\begin{figure}[t!]\centering
\includegraphics[scale=0.7]{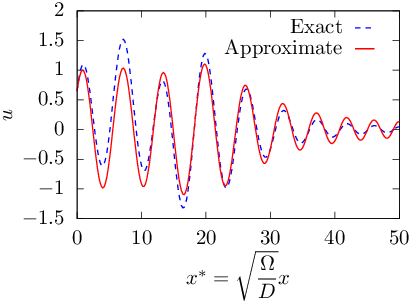}
\caption{
Shape of $u(x,t)$ for $\omega(k)=Dk^2$ at $t^\ast=\Omega t=15$.
The solid and dashed curves show the approximate solution~\eqref{eq:approximate_Dk2} and the exact solution~\eqref{eq:solution_Dk2}, respectively.
}
\label{fig5}
\end{figure}

Next, for $\omega(k)=Dk^2$, namely $v_\mathrm{g}=2DK$ and $\gamma=D$, Eq.~\eqref{eq:approximate} becomes
\begin{align}
u(x,t)&=\frac{A}{2}\left[1+C\left(\frac{2DKt-x}{\sqrt{2\pi Dt}}\right)+S\left(\frac{2DKt-x}{\sqrt{2\pi Dt}}\right)\right]\sin(\Omega t-Kx)\nonumber\\
&\quad+\frac{A}{2}\left[C\left(\frac{2DKt-x}{\sqrt{2\pi Dt}}\right)-S\left(\frac{2DKt-x}{\sqrt{2\pi Dt}}\right)\right]\cos(\Omega t-Kx)\nonumber\\
&\quad+\frac{A}{2}\left[1-C\left(\frac{2DKt+x}{\sqrt{2\pi Dt}}\right)-S\left(\frac{2DKt+x}{\sqrt{2\pi Dt}}\right)\right]\sin(\Omega t+Kx)\nonumber\\
&\quad-\frac{A}{2}\left[C\left(\frac{2DKt+x}{\sqrt{2\pi Dt}}\right)-S\left(\frac{2DKt+x}{\sqrt{2\pi Dt}}\right)\right]\cos(\Omega t+Kx).
\label{eq:approximate_Dk2}
\end{align}
This result is similar to the exact solution~\eqref{eq:solution_Dk2}; however, the $A[C(x/\sqrt{2\pi Dt})-S(x/\sqrt{2\pi Dt})]$ term does not appear.
Figure~\ref{fig5} shows this approximate solution and the exact solution~\eqref{eq:solution_Dk2} at $t^\ast=\Omega t=15$, with solid and dashed curves, respectively.
The dashed curve is the same as the solid curve in Fig.~\ref{fig2}(a).
The overall oscillating structure can be observed in the two curves, with slight differences in values.

The last example for which the approximate solution~\eqref{eq:approximate} is calculated is
\begin{equation}
\frac{\partial^2 u}{\partial t^2}=c^2\frac{\partial^2 u}{\partial x^2}-\omega_0^2u.
\label{eq:KleinGordon}
\end{equation}
This equation is the Klein--Gordon equation~\cite{Greiner} in one-dimensional space.
In the context of the Klein--Gordon equation for quantum fields, $c$ represents the speed of light and $\omega_0=mc^2/\hbar$, with $m$ representing the particle mass and $\hbar$ representing the Dirac constant.
The dispersion relation of Eq.~\eqref{eq:KleinGordon} is expressed as
\[
\omega(k)=\sqrt{c^2k^2+\omega_0^2},
\]
and $v_\mathrm{g}=c^2K/\Omega$ and $\gamma=c^2\omega_0^2/(2\Omega^3)$.
The dispersion relation in this form is also known as the Langmuir wave dispersion relation in plasma physics~\cite{Gurnett}.
Owing to the complex form of the dispersion relation, it appears that the integral in Eq.~\eqref{eq:general_solution} for this $\omega(k)$ cannot be calculated using known special functions.

\begin{figure}[t!]\centering
\includegraphics[scale=0.7]{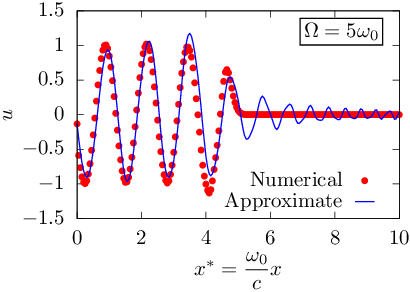}
\caption{
Numerical (points) and approximate (curve) solutions of the Klein--Gordon equation~\eqref{eq:KleinGordon} for $\Omega^\ast=\Omega/\omega_0=5$ at $t^\ast=\omega_0 t=5$.
}
\label{fig6}
\end{figure}

The model parameters $c$ and $\omega_0$ are employed to form dimensionless space and time: $x^\ast=(\omega_0/c)x$ and $t^\ast=\omega_0t$, respectively.
The angular frequency $\Omega$ or its dimensionless form $\Omega^\ast=\Omega/\omega_0$ of the wave source is regarded as a control parameter in this example.
Figure~\ref{fig6} shows the approximate solution (solid curve) and the finite element solution (circles) of the Klein--Gordon equation~\eqref{eq:KleinGordon} at $t^\ast=5$ for $\Omega^\ast=5$.
The approximation error appears to be large near $x^\ast\approx5$, the boundary between the ON and OFF states.

The approximate solution~\eqref{eq:approximate} is similar to Eq.~\eqref{eq:solution_Dk2}.
Therefore, as shown in the previous section, the ON--OFF signal propagates with group velocity $v_\mathrm{g}$, and the transition layer between the ON and OFF states is stretched at a rate proportional to $(\gamma t)^{1/2}$.
The exponent $1/2$ is suggested to be universal for characterizing the rate at which the ON--OFF signal boundary becomes indistinct owing to dispersion.

\section{Switching from the ON to the OFF state}
Here, the solution $u(x,t)$ is calculated under the opposite situation as in previous sections, in which the wave source at $x=0$ sends out the sine wave $A\sin(\Omega t-Kx)$in $t<0$, and the oscillation stops at $t=0$.
The initial and boundary conditions become
\begin{align*}
u(x,0)&=-A\sin(Kx) \quad(x>0),\\
\frac{\partial u}{\partial t}(x, 0)&=A\Omega\cos(Kx) \quad(x>0),\\
u(0, t)&=0 \quad(t\ge0).
\end{align*}
Solution $u(x,t)$ for this case is constructed by changing the plus and minus signs in Eq.~\eqref{eq:general_solution}:
\[
u(x,t)=\frac{A}{2}\sin(\Omega t-Kx)-\frac{A}{2}\sin(\Omega t+Kx)+\frac{A}{\pi}\Omega\pv{\int_0^\infty \frac{k}{k^2-K^2}\frac{1}{\omega(k)}[\cos(\omega(k)t-kx)-\cos(\omega(k)t+kx)]dk}\quad (t\ge0).
\]
Note that this solution can be expressed as sine wave $A\sin(\Omega t-Kx)$ minus Eq.~\eqref{eq:general_solution}.

\begin{figure}[t!]\centering
\raisebox{33mm}{(a)}
\includegraphics[scale=0.7]{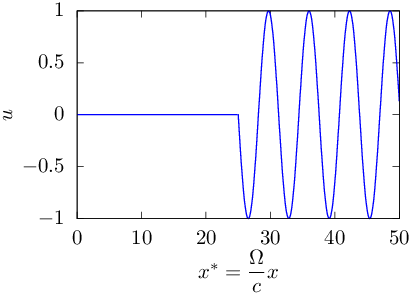}
\hspace{5mm}
\raisebox{33mm}{(b)}
\includegraphics[scale=0.7]{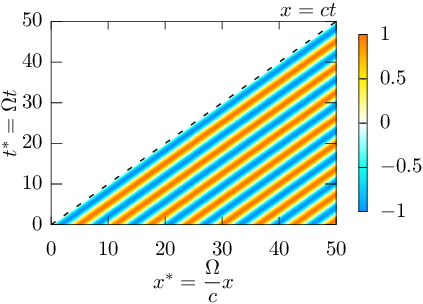}
\caption{
Signal propagation (from the ON to the OFF state) for $\omega(k)=ck$.
(a) Snapshot of $u$ at $t^\ast=25$.
(b) Spatiotemporal diagram of $u$.
The boundary between the ON and OFF states travels with velocity $c$, represented by the dashed diagonal line.
}
\label{fig7}
\end{figure}

For the nondispersive case $\omega(k)=ck$,
\[
u(x, t)=\frac{A}{2}(1-\sgn(ct-x))\sin(\Omega t-Kx)=
\begin{cases}
A\sin(\Omega t-Kx) & (x\ge ct),\\
0 & (x<ct)
\end{cases}
\]
is obtained.
Figure~\ref{fig7} shows this solution.
The clear boundary between the ON and OFF states moves with velocity $c$.

For the dispersion relation $\omega(k)=Dk^2$, $u(x,t)$ becomes
\begin{align}
u(x, t)&=\frac{A}{2}\left[1-C\left(\frac{2DKt-x}{\sqrt{2\pi Dt}}\right)-S\left(\frac{2DKt-x}{\sqrt{2\pi Dt}}\right)\right]\sin(\Omega t-Kx)\nonumber\\
&\quad-\frac{A}{2}\left[C\left(\frac{2DKt-x}{\sqrt{2\pi Dt}}\right)-S\left(\frac{2DKt-x}{\sqrt{2\pi Dt}}\right)\right]\cos(\Omega t-Kx)\nonumber\\
&\quad-\frac{A}{2}\left[1-C\left(\frac{2DKt+x}{\sqrt{2\pi Dt}}\right)-S\left(\frac{2DKt+x}{\sqrt{2\pi Dt}}\right)\right]\sin(\Omega t+Kx)\nonumber\\
&\quad+\frac{A}{2}\left[C\left(\frac{2DKt+x}{\sqrt{2\pi Dt}}\right)-S\left(\frac{2DKt+x}{\sqrt{2\pi Dt}}\right)\right]\cos(\Omega t+Kx)\nonumber\\
&\quad-A\left[C\left(\frac{x}{\sqrt{2\pi Dt}}\right)-S\left(\frac{x}{\sqrt{2\pi Dt}}\right)\right].
\label{eq:solution_Dk2_ONtoOFF}
\end{align}
Taking the $t\to\infty$ limit and using $C(\infty)=S(\infty)=1/2$, we immediately obtain
\[
\lim_{t\to\infty}u(x,t)=0
\]
for any finite $x$.
Thus, sufficiently long after the ON signal passes, the complete OFF state is realized for any $x$.

\begin{figure}\centering
\raisebox{34mm}{(a)}
\includegraphics[scale=0.7]{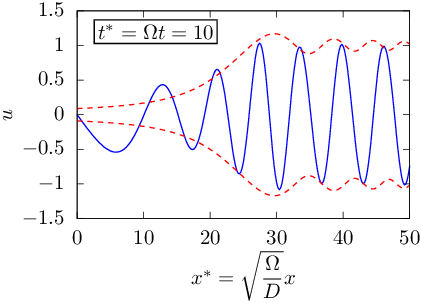}
\hspace{3mm}
\raisebox{34mm}{(b)}
\includegraphics[scale=0.7]{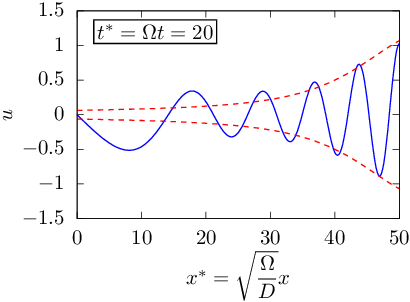}
\hspace{3mm}
\raisebox{34mm}{(c)}
\includegraphics[scale=0.7]{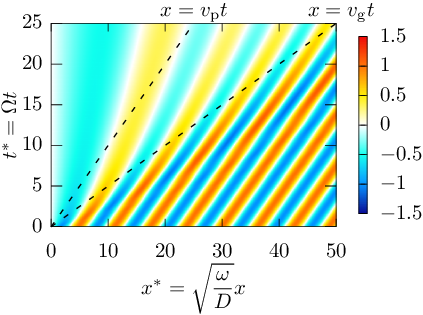}
\caption{
Solution $u(x,t)$ given by Eq.~\eqref{eq:solution_Dk2_ONtoOFF} with a solid curve, and approximate envelope $\tilde{A}(x,t)$ given by Eq.~\eqref{eq:approximate_envelope_ONtoOFF} with a dashed curve at $t^\ast=10$ (a) and $20$ (b).
(c) Spatiotemporal diagram of $u$.
}
\label{fig8}
\end{figure}

As in Section~\ref{sec4}, the approximate envelope of Eq.~\eqref{eq:solution_Dk2_ONtoOFF} can be obtained as
\begin{equation}
\tilde{A}(x,t)=\frac{A}{\sqrt{2}}\sqrt{\left[C\left(\frac{2DKt-x}{\sqrt{2\pi Dt}}\right)-\frac{1}{2}\right]^2+\left[S\left(\frac{2DKt-x}{\sqrt{2\pi Dt}}\right)-\frac{1}{2}\right]^2}.
\label{eq:approximate_envelope_ONtoOFF}
\end{equation}
The graphs of $u(x, t)$ and $\tilde{A}(x,t)$ at $t^\ast=10$ and $20$ are shown in Figs.~\ref{fig8}(a) and (b), respectively.
Figure~\ref{fig8}(c) illustrates the spatiotemporal property of $u(x,t)$.
As in the above figures, the ON--OFF signal, i.e., the boundary between the ON and the OFF states, moves with the group velocity.

The slope of the tangent line of $u(x,t)$ at $x=0$ is calculated as follows:
\[
\frac{\partial u}{\partial x}(0, t)=-AK\left[\left(1-C\left(\sqrt{\frac{2\Omega t}{\pi}}\right)-S\left(\sqrt{\frac{2\Omega t}{\pi}}\right)\right)\cos\Omega t+\left(C\left(\sqrt{\frac{2\Omega t}{\pi}}\right)-S\left(\sqrt{\frac{2\Omega t}{\pi}}\right)\right)\sin\Omega t\right].
\]
For a sufficiently large $t$, using the asymptotic expansions~\eqref{eq:asymptotic},
\[
\frac{\partial u}{\partial x}(0, t)\simeq-\frac{A}{\sqrt{2\pi Dt}}
\]
is obtained.
Therefore, the slope converges to $0$ at the rate asymptotically proportional to $t^{-1/2}$.

\section{Higher-complexity ON--OFF switching}\label{sec7}
A higher-complexity ON--OFF switching pattern of the wave source at $x=0$ is explored.
The wave source does not oscillate until $t=0$; it starts oscillating at $t=0$ and stops oscillating after $n$ cycles of oscillation.
The corresponding solution is denoted as $u_n(x, t)$.
By setting the oscillation period as $T=2\pi/\Omega$, the oscillation pattern at $x=0$ is expressed as
\begin{equation}
u_n(0, t)=
\begin{cases}
0 & (t<0),\\
A\sin\Omega t & (0\le t\le nT),\\
0 & (t>nT).
\end{cases}
\label{eq:u_n0}
\end{equation}

The simple switching from OFF to ON at $t=0$ considered in Sections~\ref{sec2}--\ref{sec5} can be formally expressed as $u_\infty(x, t)$.
Thus,
\[
u_\infty(0, t)=
\begin{cases}
0 & (t<0),\\
A\sin\Omega t & (t\ge0).
\end{cases}
\]
A key relation for finding $u_n(x,t)$ is $u_n(0, t)=u_\infty(0, t)-u_\infty(0, t-nT)$.
Owing to linearity, the following solution is obtained:
\[
u_n(x,t) = u_\infty(x,t)-u_\infty(x, t-nT).
\]
The specific formula is obtained by replacing $u_\infty$ with $u$ in Eq.~\eqref{eq:general_solution} for the general dispersion relation, in Eq.~\eqref{eq:solution_ck} for $\omega=ck$, and in Eq.~\eqref{eq:solution_Dk2} for $\omega=Dk^2$.

\begin{figure}[t!]\centering
\raisebox{34mm}{(a)}
\includegraphics[scale=0.7]{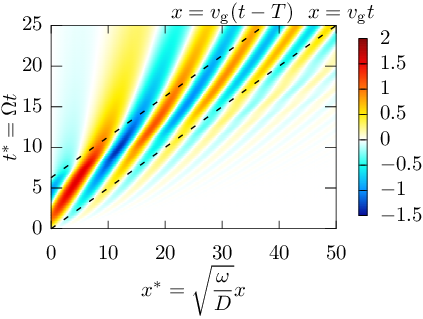}
\raisebox{34mm}{(b)}
\includegraphics[scale=0.7]{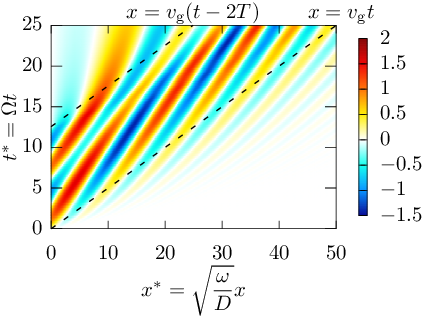}
\raisebox{34mm}{(c)}
\includegraphics[scale=0.7]{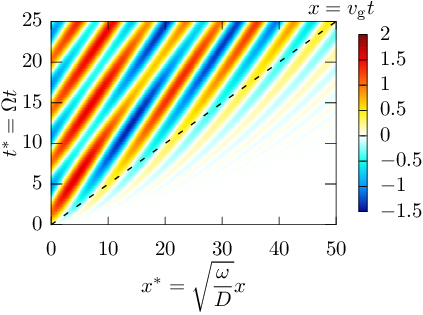}
\caption{
Spatiotemporal diagram of $u_n(x,t)$ for $n=1$ (a), $2$ (b), and $5$ (c), under dispersion relation $\omega=Dk^2$.
The dashed diagonal lines indicate $x=v_\mathrm{g}t$ and $x=v_\mathrm{g}(t-nT)$.
}
\label{fig9}
\end{figure}

For $\omega=ck$, $u_n(x, t)$ indicates that a sine curve with $n$ cycles moves with velocity $c$, maintaining its shape, which is d'Alembert's (weak) solution.
The spatiotemporal diagrams of $u_n$ for $\omega=Dk^2$ with $n=1$, $2$, and $5$ are shown in Fig.~\ref{fig9}, indicating that the ON--OFF signals travel with the group velocity.
Meanwhile, owing to dispersion, the wave is not perfectly periodic and, it broadens and flattens with time.

\section{Concluding remarks and discussion}
In this study, the propagation of ON--OFF wave signals is examined.
The integral-form solution~\eqref{eq:general_solution} valid for the general dispersion relation $\omega=\omega(k)$ is derived, and the exact solution in a closed form is obtained for dispersion relations $\omega(k)=ck$ and $\omega(k)=Dk^2$.
In particular, the solution for $\omega(k)=Dk^2$, given in Eq.~\eqref{eq:solution_Dk2}, is complex and nontrivial.
There may be other dispersion relations that provide the closed form of $u$, and the search for such dispersion relations will be performed in a future study.

To analyze solution $u(x,t)$ in Eq.~\eqref{eq:solution_Dk2} for $\omega(k)=Dk^2$, approximate envelope $\tilde{A}(x,t)$ was calculated.
The analysis of $\tilde{A}(x,t)$ revealed that the boundary between the ON and OFF states moves with group velocity $v_\mathrm{g}$, and the boundary thickness increases at $t^{1/2}$ with time, owing to the dispersion.
From the analysis of the approximate solution in Section~\ref{sec5}, this exponent $1/2$ was considered to be universal whenever $\left.d^2\omega(k)/dk^2\right|_{k=K}\ne0$.
Determining the exact envelope is a challenging task, which will provide detailed information on the properties of the ON--OFF signal propagation.

Although an approximate solution~\eqref{eq:approximate} was found, the accuracy of the approximation was not ascertained.
For example, in Fig.~\ref{fig6} showing the results for the Klein--Gordon equation~\eqref{eq:KleinGordon}, finite element simulation (circles) provides $u(x,t)\approx0$ for $x^\ast\gtrsim6$ at $t^\ast=5$.
In contrast, the approximate solution (solid curve) displays a wave shape with a slowly decreasing amplitude.
This discrepancy can be explained by evaluating the accuracy of the approximation.
Although improving the accuracy is important, it is difficult to achieve.
Since the integrals must be calculated after the Taylor expansion of the dispersion relation, the order of the approximation cannot be freely increased.

Extending the idea in Section~\ref{sec7}, solution $u(x, t)$ is calculated for the higher-complexity ON--OFF switching scenario, such as the transmission of Morse code.
Furthermore, by applying Fourier decomposition to the oscillation pattern of the wave source, a propagation solution in which the oscillation of the wave source is not harmonic is constructed.

The solution~\eqref{eq:general_solution} is expressed as a superposition of sine waves.
Since linearity is assumed, the method in this study is not directly applicable to nonlinear waves, such as solitons.
Meanwhile, the proposed method is not restricted to a specific wave.
Solving the initial-boundary value problem without specifying the governing partial differential equation would be an interesting approach.

If the dispersion relation $\omega(k)$ is not simple, the integral in Eq.~\eqref{eq:general_solution} cannot be expressed using existing mathematical functions.
Even in such a case, we expect that a numerical solution is obtained by evaluating the integral.
Although the integral is the Cauchy principal-value integral, numerical procedures to evaluate integrals of this type have been developed~\cite{Klerk}.
The advantage of this method is that we do not have to discretize the governing partial differential equation.
The value of $u(x,t)$ at arbitrary position $x$ and time $t$ will be computed without generating a spatial grid and iterating time steps.
Hence, the method of evaluating Eq.~\eqref{eq:general_solution} will be suitable for analyzing the wave at a point $x$ far distant from the wave source $x=0$ or asymptotic behavior at sufficiently large $t$.

Since the solution~\eqref{eq:solution_Dk2} in this study is for the Euler--Bernoulli beam equation~\eqref{eq:EulerBernoulli}, this study is directly connected to the analysis of vibrating beams and elastic waves that is fundamental in vibration engineering~\cite{Genta}.
The author believes that this study can be applied to engineering problems, such as the vibration of machine and building components.
In a more general framework, by precisely investigating the mathematical mechanism behind why the exact solution can be obtained, the theoretical understanding of wave propagation will improve, and this development will give insights to complex wave phenomena.

\section*{Acknowledgment}
This study was partially supported by a Grant-in-Aid for Scientific Research (C) 19K03656 from Japan Society for the Promotion of Science.

\appendix
\section{Proof of integrals used in the main part}
Here we briefly derive integrals that appear in the main part.

\subsection{$\cos(Kx)\sgn(x)=\dfrac{2}{\pi}\pv{\displaystyle\int_0^\infty \dfrac{k}{k^2-K^2}\sin(kx)dk}$}\label{apdx:a2}
Using the integral shown in the above subsection, we have
\begin{align*}
\cos(Kx)\sgn(x)&=\frac{2}{\pi}\int_0^\infty \frac{1}{k}\sin(kx)\cos(Kx)dk\\
&=\frac{1}{\pi}\lim_{\varepsilon\to0+}\int_\varepsilon^\infty \frac{1}{k}[\sin((k+K)x)+\sin((k-K)x)]dk\\
&=\frac{1}{\pi}\lim_{\varepsilon\to0+}\left[\int_{K+\varepsilon}^\infty \frac{\sin(kx)}{k-K}dk+\int_{-K+\varepsilon}^\infty \frac{\sin(kx)}{k+K}dk\right]\\
&=\frac{1}{\pi}\lim_{\varepsilon\to0+}\left[\int_{K+\varepsilon}^\infty \frac{2k}{k^2-K^2}\sin(kx)dk+\int_{-K+\varepsilon}^{K+\varepsilon}\frac{\sin(kx)}{k+K}dk\right]
\end{align*}
$k=0$ is a removable singularity of the integrand in the first line; in fact, the integrand converges to $x\cos(Kx)$ as $k\to0$.
Nevertheless, we introduce the limit $\varepsilon\to0+$ explicitly so that the singularity can be correctly tracked in the calculation below.
The second integral in the last line can be further calculated to
\begin{align*}
\int_{-K+\varepsilon}^{K+\varepsilon}\frac{\sin(kx)}{k+K}dk&=\left(\int_{-K+\varepsilon}^0+\int_0^{K+\varepsilon}\right)\frac{\sin(kx)}{k+K}dk\\
&=\int_0^{K-\varepsilon}\frac{\sin(kx)}{k-K}dk+\int_0^{K-\varepsilon}\frac{\sin(kx)}{k+K}dk + \int_{K-\varepsilon}^{K+\varepsilon}\frac{\sin(kx)}{k+K}dk\\
&=\int_0^{K-\varepsilon}\frac{2k}{k^2-K^2}\sin(kx)dk+ \int_{K-\varepsilon}^{K+\varepsilon}\frac{\sin(kx)}{k+K}dk.
\end{align*}
The second integral vanishes as $\varepsilon\to0$, because its integrand $\sin(kx)/(k+K)$ is finite at $k=K$.
Therefore, we obtain
\begin{align*}
\cos(Kx)\sgn(x)&=\frac{2}{\pi}\lim_{\varepsilon\to0+}\left[\int_0^{K-\varepsilon}\frac{k}{k^2-K^2}\sin(kx)dk+\int_{K+\varepsilon}^\infty \frac{k}{k^2-K^2}\sin(kx)dk\right]\\
&=\frac{2}{\pi}\pv{\int_0^\infty \frac{k}{k^2-K^2}\sin(kx)dk}.
\end{align*}

\subsection{$\pv\displaystyle\int_0^\infty\dfrac{1}{k^2-K^2}\cos(\alpha k)dk=-\dfrac{\pi\sin(\alpha K)}{2K}\sgn(\alpha)$}\label{apdx:a3}
Similar to the above appendix,
\begin{align*}
-\frac{\pi\sin(\alpha K)}{2K}\sgn(\alpha)&=-\frac{\pi}{2K}\frac{2}{\pi}\int_0^\infty \frac{1}{k}\sin(\alpha k)\sin(\alpha K)dk\\
&=\lim_{\varepsilon\to0+}\left[\int_{K+\varepsilon}^\infty \frac{\cos(\alpha k)}{k^2-K^2}dk-\frac{1}{2K}\int_{-K+\varepsilon}^{K+\varepsilon} \frac{\cos(\alpha k)}{k+K}dk\right].
\end{align*}
The second integral becomes
\[
-\frac{1}{2K}\int_{-K+\varepsilon}^{K+\varepsilon} \frac{\cos(\alpha k)}{k+K}dk
=\int_0^{K-\varepsilon}\frac{\cos(\alpha k)}{k^2-K^2}dk-\frac{1}{2K}\int_{K-\varepsilon}^{K+\varepsilon}\frac{\cos(\alpha k)}{k+K}dk.
\]
Since the last integral vanishes as in the above Appendix, we have
\[
-\frac{\pi\sin(\alpha K)}{2K}\sgn(\alpha)=\lim_{\varepsilon\to0+}\left(\int_0^{K-\varepsilon}+\int_{K+\varepsilon}^\infty\right) \frac{\cos(\alpha k)}{k^2-K^2}dk
=\pv{\displaystyle\int_0^\infty\dfrac{1}{k^2-K^2}\cos(\alpha k)dk}.
\]

\section{Derivation of exact solution \eqref{eq:solution_Dk2} for $\omega(k)=Dk^2$}\label{apdx:b}
The following integral is calculated:
\[
u(x, t)=\frac{A}{2}\sin(\Omega t-Kx)+\frac{A}{2}\sin(\Omega t+Kx)-\frac{A\Omega}{\pi D}\pv{\int_0^\infty \frac{1}{k^2-K^2}\frac{1}{k}[\cos(Dk^2t-kx)-\cos(Dk^2t+kx)]dk}.
\]
Using 
\[
\frac{1}{k^2-K^2}\frac{1}{k}=\frac{1}{2K^2}\frac{1}{k-K}+\frac{1}{2K^2}\frac{1}{k+K}-\frac{1}{K^2}\frac{1}{k}
\]
and $\Omega=DK^2$, the integral term becomes
\begin{align*}
&-\frac{A}{2\pi}\pv{\int_{-K}^\infty \frac{1}{k}[\cos(D(k+K)^2t-(k+K)x)-\cos(D(k+K)^2t+(k+K)x)]dk}\\
&-\frac{A}{2\pi}\int_K^\infty \frac{1}{k}[\cos(D(k-K)^2t-(k-K)x)-\cos(D(k-K)^2t+(k-K)x)]dk\\
&+\frac{A}{\pi}\int_0^\infty \frac{1}{k}[\cos(Dk^2t-kx)-\cos(Dk^2t+kx)]dk\\
&\eqqcolon I_1(x,t)+I_2(x,t)+I_3(x,t).
\end{align*}
By the definition of the Cauchy principal value,
\[
I_1(x,t)=-\frac{A}{2\pi}\lim_{\varepsilon\to0+}\left(\int_\varepsilon^\infty+\int_{-K}^{-\varepsilon}\right) \frac{1}{k}[\cos(D(k+K)^2t-(k+K)x)-\cos(D(k+K)^2t+(k+K)x)]dk.
\]
As shown below, these two integrals exist independently as $\varepsilon\to0$.
The second integral can be written as
\begin{align*}
&-\frac{A}{2\pi}\int_{-K}^0 \frac{1}{k}[\cos(D(k+K)^2t-(k+K)x)-\cos(D(k+K)^2t+(k+K)x)]dk\\
&=\frac{A}{2\pi}\int_0^K \frac{1}{k}[\cos(D(-k+K)^2t-(-k+K)x)-\cos(D(-k+K)^2t+(-k+K)x)]dk\\
&=-\frac{A}{2\pi}\int_0^K \frac{1}{k}[\cos(D(k-K)^2-(k-K)x)-\cos(D(k-K)^2+(k-K)x)]dk.
\end{align*}
Therefore,
\begin{align*}
I_1(x,t)+I_2(x,t)&=-\frac{A}{2\pi}\int_0^\infty \frac{1}{k}[\cos(D(k+K)^2t-(k+K)x)-\cos(D(k+K)^2t+(k+K)x)]dk\\
&\quad-\frac{A}{2\pi}\int_0^\infty \frac{1}{k}[\cos(D(k-K)^2t-(k-K)x)-\cos(D(k-K)^2t+(k-K)x)]dk\\
&=\frac{A}{\pi}\int_0^\infty\frac{1}{k}[\cos(Dk^2t)\sin((2DKt-x)k)\sin(\Omega t-Kx)+\sin(Dk^2t)\sin((2DKt-x)k)\cos(\Omega t-Kx)\\
&\qquad -\cos(Dk^2t)\sin((2DKt+x)k)\sin(\Omega t+Kx) -\sin(Dk^2t)\sin((2DKt+x)k)\cos(\Omega t+Kx)]dk.
\end{align*}

The closed-form solution of $u(x,t)$ is obtained by applying the formulas~\cite{Oberhettinger}
\begin{subequations}\label{eq:B1}
\begin{align}
\int_0^\infty \frac{1}{k}\sin(\alpha k^2)\sin(\beta k)dk&=\frac{\pi}{2}\sgn(\alpha)\left[C\left(\frac{\beta}{\sqrt{2\pi|\alpha|}}\right)-S\left(\frac{\beta}{\sqrt{2\pi|\alpha|}}\right)\right],\\
\int_0^\infty \frac{1}{k}\cos(\alpha k^2)\sin(\beta k)dk&=\frac{\pi}{2}\left[C\left(\frac{\beta}{\sqrt{2\pi|\alpha|}}\right)+S\left(\frac{\beta}{\sqrt{2\pi|\alpha|}}\right)\right],
\end{align}
\end{subequations}
to each term of $I_1(x, t)+I_2(x,t)$ and $I_3(x,t)$.
These formulas can be regarded as the Fourier sine transforms of $\sin(\alpha k^2)/k$ and $\cos(\alpha k^2)/k$, respectively.

\section{Calculation of the approximate solution~\eqref{eq:approximate}}\label{apdx:c}
Here, the derivation of the approximate solution~\eqref{eq:approximate} is outlined.
The integral in Eq.~\eqref{eq:general_solution} is calculated as
\begin{align*}
&\pv{\int_0^\infty\frac{k}{k^2-K^2}\frac{1}{\omega(k)}[\cos(\omega(k)t-kx)-\cos(\omega(k)t+kx)]dk}\nonumber\\
&=\frac{1}{2}\pv{\int_0^\infty\left(\frac{1}{k-K}+\frac{1}{k+K}\right)\frac{1}{\omega(k)}[\cos(\omega(k)t-kx)-\cos(\omega(k)t+kx)]dk}\nonumber\\
&=\frac{1}{2}\pv{\int_{-K}^\infty\frac{1}{k}\frac{1}{\omega(k+K)}[\cos(\omega(k+K)t-(k+K)x)-\cos(\omega(k+K)t+(k+K)x)]dk}\nonumber\\
&\quad+\frac{1}{2}\int_K^\infty\frac{1}{k}\frac{1}{\omega(k-K)}[\cos(\omega(k-K)t-(k-K)x)-\cos(\omega(k-K)t+(k-K)x)]dk.
\end{align*}
As in Appendix~\ref{apdx:b}, the Cauchy principal-value integral is split into
\begin{align*}
&\frac{1}{2}\pv{\int_{-K}^\infty\frac{1}{k}\frac{1}{\omega(k+K)}[\cos(\omega(k+K)t-(k+K)x)-\cos(\omega(k+K)t+(k+K)x)]dk}\\
&=\frac{1}{2}\left(\int_0^\infty+\int_{-K}^0\right)\frac{1}{k}\frac{1}{\omega(k+K)}[\cos(\omega(k+K)t-(k+K)x)-\cos(\omega(k+K)t+(k+K)x)]dk.
\end{align*}
The integral from $-K$ to $0$ can be manipulated as follows:
\begin{align*}
&\frac{1}{2}\int_{-K}^0 \frac{1}{k}\frac{1}{\omega(k+K)}[\cos(\omega(k+K)t-(k+K)x)-\cos(\omega(k+K)t+(k+K)x)]dk\\
&=\frac{1}{2}\int_0^K \frac{1}{k}\frac{1}{\omega(-k+K)}[\cos(\omega(-k+K)t-(k-K)x)-\cos(\omega(-k+K)t+(k-K)x)]dk.
\end{align*}
Here, it is assumed that the dispersion relation $\omega(k)$ is an even function: $\omega(k)=\omega(-k)$.
This is a natural assumption, which reflects that the wave medium has spatial inversion symmetry; this assumption often holds for real dispersion relations.
Thus,
\begin{align}
u(x,t)&=\frac{A}{2}\sin(\Omega t-Kx)+\frac{A}{2}\sin(\Omega t+Kx)\nonumber\\
&\quad+\frac{A\Omega}{2\pi}\int_0^\infty\frac{1}{k}\frac{1}{\omega(k+K)}[\cos(\omega(k+K)t-(k+K)x)-\cos(\omega(k+K)t+(k+K)x)]dk\nonumber\\
&\quad+\frac{A\Omega}{2\pi}\int_0^\infty\frac{1}{k}\frac{1}{\omega(k-K)}[\cos(\omega(k-K)t-(k-K)x)-\cos(\omega(k-K)t+(k-K)x)]dk.
\label{eqC1}
\end{align}
This expression is still exact.

The Taylor expansion of $\omega(k)$ around $k=K$ and $k=-K$ is adopted for the first and second integrals, respectively:
\[
\omega(k+K)=\Omega+v_\mathrm{g} k+\gamma k^2+O(k^3),\quad
\omega(k-K)=\Omega-v_\mathrm{g} k+\gamma k^2+O(k^3),
\]
where
\[
\gamma=\frac{1}{2}\left.\frac{d^2\omega(k)}{dk^2}\right|_{k=K}.
\]
The coefficients of $k^2$ in both expansions are identical ($\gamma$), owing to the assumption that $\omega(k)=\omega(-k)$.

By substituting the above expansions for $\omega(k\pm K)$ in the cosine functions and using $\omega(k\pm K)\simeq\Omega$ for $1/\omega(k\pm K)$ factors, Eq.~\eqref{eqC1} is approximated to yield
\begin{align*}
u(x,t)&\simeq\frac{A}{2}\sin(\Omega t-Kx)+\frac{A}{2}\sin(\Omega t+Kx)\\
&\quad+\frac{A}{\pi}\int_0^\infty \frac{1}{k}[-\sin(\gamma k^2t)\sin((v_\mathrm{g}t-x)k)\cos(\Omega t-Kx)-\cos(\gamma k^2t)\sin((v_\mathrm{g}t-x)k)\sin(\Omega t-Kx)\\
&\qquad\qquad +\sin(\gamma k^2t)\sin((v_\mathrm{g}t+x)k)\cos(\Omega t+Kx)+\cos(\gamma k^2t)\sin((v_\mathrm{g}t+x)k)\sin(\Omega t+Kx)]dk.
\end{align*}
The final result~\eqref{eq:approximate} is obtained using Eq.~\eqref{eq:B1}.


\begin{thebibliography}{99}
\bibitem{Crawford}
Crawford~Jr~F~S 1965 \textit{Berkeley Physics Course: Waves} (NY: McGraw-Hill)

\bibitem{Lighthill}
Lighthill~J 1978 \textit{Waves in Fluids} (Cambridge: Cambridge University Press)

\bibitem{Nettel}
Nettel~S 2009 Wave Physics (Berlin: Springer)

\bibitem{Narayanan}
Narayanan~A~S and Saha~S~K 2015 \textit{Waves and Oscillations in Nature} (Boca Raton, FL: CRC Press)

\bibitem{Phillips}
Phillips~A~C 2003 \textit{Introduction to Quantum Mechanics} (Chichester: Wiley)

\bibitem{Creighton}
Creighton~J~D~E and Anderson~W~G 2011 \textit{Gravitational-Wave Physics and Astronomy} (Weinheim: Wiley)

\bibitem{Tse}
Tse~D and Viswanath~P 2005 \textit{Fundamentals of Wireless Communication} (Cambridge: Cambridge University Press)

\bibitem{Pierce}
Pierce~A~D 2019 \textit{Acoustics: An Introduction to Its Physical Principles and Applications} (Cham: Springer)

\bibitem{Taranath}
Taranath~B~S 2004 \textit{Wind and Earthquake Resistant Buildings} (Boca Raton, FL: CRC Press)

\bibitem{Holton}
Holton~J~R and Hakim~G~J 2013 \textit{An Introduction to Dynamic Meteorology} (Waltham: Academic Press)

\bibitem{Boyd}
Boyd~R~W and Gauthier~D~J 2002 ``Slow'' and ``fast'' light \textit{Progress in Optics} vol 43 ed E~Wolf (Amsterdam: Elsevier) p 497

\bibitem{Sommerfeld}
Sommerfeld~A 1964 \textit{Lectures on Theoretical Physics} vol 4 (NY: Academic Press)

\bibitem{Brillouin}
Brillouin~L 1960 \textit{Wave Propagation and Group Velocity} (NY: Academic Press)

\bibitem{Oughstun}
Oughstun~K~E and Sherman~G~C 1994 \textit{Electromagnetic Pulse Propagation in Causal Dielectrics} (Berlin: Springer)

\bibitem{Jeong}
Jeong~H, Dawes~A~M~C and Gauthier~D~J 2006 Direct observation of optical precursors in a region of anomalous dispersion \textit{Phys.\ Rev.\ Lett.} \textbf{96} 143901

\bibitem{Oughstun2005}
Oughstun~K~E 2005 Dynamical evolution of the Brillouin precursor in Rocard--Powles--Debye model dielectrics \textit{IEEE Trans. Antennas and Propagation} \textbf{53} 1582--90

\bibitem{Alejos}
Alejos~A~V, Dawood~M and Medina~L 2011 Experimental dynamical evolution of the Brillouin precursor for broadband wireless communication through vegetation \textit{Prog.\ Electromagnetics Res.} \textbf{111} 291--309

\bibitem{Cartwright}
Cartwright~N~A and Muller~K 2023 Precursors for synthetic aperture radar \textit{Inverse Problems} \textbf{39} 064003

\bibitem{Ahlfors}
Ahlfors~L~V 1979 \textit{Complex Analysis} (NY: McGraw-Hill)

\bibitem{Bartle}
Bartle~R~G 1996 Return to the Riemann integral \textit{Amer.\ Math.\ Monthly} \textbf{103} 625--32

\bibitem{Thorne}
Thorne~K~S and Blandford~R~D 2017 \textit{Modern Classical Physics} (Princeton, NJ: Princeton University Press)

\bibitem{Cain}
Cain~G and Meyer~G~H 2006 \textit{Separation of Variables for Partial Differential Equations} (Milton Park: Taylor and Francis)

\bibitem{Hagedorn}
Hagedorn~P and DasGupta~A 2007 \textit{Vibrations and Waves in Continuous Mechanical Systems} (NY: Wiley)

\bibitem{Olver}
Olver~F~W, Lozier~D~W, Boisvert~R~F and Clark~C~W 2010 \textit{NIST Handbook of Mathematical Functions} (Cambridge: Cambridge University Press)

\bibitem{Kikuchi}
Kikuchi~N 1986 \textit{Finite Element Methods in Mechanics} (Cambridge: Cambridge University Press)

\bibitem{Greiner}
Greiner~W 2000 \textit{Relativistic Quantum Mechanics} (Berlin: Springer)

\bibitem{Gurnett}
Gurnett~D~A and Bhattacharjee~A 2017 \textit{Introduction to Plasma Physics} (Cambridge: Cambridge University Press)

\bibitem{Klerk}
de Klerk~J~H 2011 \textit{AIP Conf Proc.} \textbf{1389} 456--9

\bibitem{Genta}
Genta~G 2009 \textit{Vibration Dynamics and Control} (NY: Springer)

\bibitem{Oberhettinger}book
Oberhettinger~F 1990 \textit{Tables of Fourier Transforms and Fourier Transforms of Distributions} (Berlin: Springer)

\end{thebibliography}
\end{document}